\documentclass[onecolumn,amsmath,amssymb]{qm2008_abs}
\usepackage{graphicx}
\usepackage{dcolumn}
\usepackage{bm}
\usepackage{subfigure}
\begin{document}

\title{{\Large Eight-quark interactions as a chiral thermometer }}

\bigskip
\bigskip
\author{\large J. Moreira, A. A. Osipov, B. Hiller, A. H. Blin, J. Providencia}
\email{jmoreira@teor.fis.uc.pt,osipov@nusun.jinr.ru ,brigitte@teor.fis.uc.pt, alex@teor.fis.uc.pt, providencia@teor.fis.uc.pt}
\affiliation{Centro de Fisica Teórica, Dep. Fisica Univ. Coimbra, Portugal}
\bigskip
\bigskip

\begin{abstract}
\leftskip1.0cm
\rightskip1.0cm

A NJL Lagrangian extended to six \cite{Hooft, Bernard, ReinhardtAlkofer} and eight quark interactions \cite{Osipov:2005b} is applied to study temperature effects \cite{Osipov:2007b} (SU(3) flavor limit, massless case), and \cite{Osipov:2007c} (realistic massive case). The transition temperature can be considerably reduced as compared to the standard approach, in accordance with recent lattice calculations \cite{latticecalc}.  The mesonic spectra built on the spontaneously broken vacuum induced by the 't Hooft interaction strength, as opposed to the
commonly considered case driven by the four-quark coupling, undergoes a 
rapid crossover to the unbroken phase, with a slope and at a
temperature which is regulated by the strength of the OZI violating 
eight-quark interactions. This strength can be adjusted in consonance 
with the four-quark coupling and leaves the spectra unchanged, except 
for the sigma meson mass, which decreases. A first order transition behavior 
is also a possible solution within the present approach.
\end{abstract}

\maketitle

\section{Introduction}

The Nambu--Jona-Lasinio Model (\cite{NJL}, for reviews see e.g.\cite{Klevansky,Kunihiro}) is an effective quark theory in which the gluonic degrees of freedom are frozen into pointlike interactions. Despite its simplicity it incorporates some of QCD's key features and has been widely used as a tool for the study of low energy hadron phenomenology. It incorporates a mechanism for dynamical chiral symmetry breaking in which the quark condensates appear as order parameters and the light mesons are regarded as Goldstone bosons. Here we present our work with a version of this model extended to include 6 and 8 quark interactions in the light quark sector (u, d and s). The inclusion of these multi-quark interactions can be motivated through an analogy with the case for QCD in the instanton gas model \cite{Simunov}. The 8 quark terms are needed to stabilize the ground state \cite{Osipov:2005b, 2}.
The two flavor NJL model with inclusion of 8q interactions
has been considered in \cite{japoneses}.
 The terms considered constitute the most general spin zero  chirally symmetric non-derivative combinations.

The model lagrangian which we consider can be decomposed in the following terms: the free Dirac Lagrangian $\mathcal{L}_D$, the Nambu--Jona-Lasinio 4-quark interaction ($\mathcal{L}_{NJL}$) which, if strong enough, results in a double well potential with the resulting nonvanishing quark condensates in the physical vacuum, the 6-quark flavour determinant 't Hooft term ($\mathcal{L}_{H}$) which is needed to break the unwanted $U_A(1)$ symmetry  and an 8 quark interaction $\mathcal{L}_{8q}=\mathcal{L}^{(1)}_{8q}+\mathcal{L}^{(2)}_{8q}$ (where $\mathcal{L}^{(1)}_{8q}$ violates the Okubo-Zweig-Iizuka (OZI) rule). Chiral symmetry is explicitly broken by the inclusion of the current quark masses (here we will consider the chiral symmetric case $m_u=m_d=m_s=0$ and realistic values in the case $m_u=m_d\neq m_s\neq 0$).

The Lagrangian is written as

\begin{align}
\mathcal{L}_{eff} & = \mathcal{L}_D +\mathcal{L}_{NJL}+\mathcal{L}_{H}+\mathcal{L}_{8q} &\qquad &\nonumber\\
\mathcal{L}_D &=\overline{q}\left(\imath\gamma^\mu \partial_\mu-\hat{m}\right)q &\qquad  
\mathcal{L}_{8q} & =\mathcal{L}^{(1)}_{8q}+\mathcal{L}^{(2)}_{8q} 			\nonumber\\
\mathcal{L}_{NJL}&=\frac{G}{2}\left[\left(\overline{q}\lambda_a q\right)^2 + \left(\overline{q}\imath \gamma5 \lambda_a q\right)^2\right] & \qquad
\mathcal{L}^{(1)}_{8q} & =8 g_1\left[\left(\overline{q}_iP_{R} q_{m}\right)\left(\overline{q}_mP_L q_i\right)\right]^2 	\nonumber\\
\mathcal{L}_{H} & = \kappa \left(\mathrm{det}\overline{q}P_Lq +\mathrm{det}\overline{q} P_R q \right)& \qquad
\mathcal{L}^{(2)}_{8q} & =16 g_2\left(\overline{q}_iP_R q_m\right)\left(\overline{q}_mP_L q_j\right)
					 \left(\overline{q}_jP_R q_K\right)\left(\overline{q}_KP_L q_i\right) \nonumber
\end{align}

\section{Evaluation of the functional integrals}
The lagrangian is bosonized using the functional identity\cite{ReinhardtAlkofer}
\begin{align}
		1&= \int \prod_a\mathcal{D}s_a\mathcal{D}p_a\mathcal{D}\sigma_a\mathcal{D}\phi_a
		e^{\imath\int d^4x[\sigma_a(s_a-\overline{q}\lambda_a q)+\phi_a(p_a-\overline{q}\imath\gamma_5\lambda_a q)]}\nonumber
\end{align}

Thus we introduce two sets of flavour nonet bosonic fields: $\{\sigma_a,\phi_a\}$, related to the physical scalar and pseudo-scalar mesons,  and $\{s_a,p_a\}$ which are auxiliary fields. Through this procedure we can separate the vacuum-to-vacuum amplitude in two parts: 
\begin{align}
  Z=&\int \prod_a \mathcal{D}\sigma_a \mathcal{D}\phi_a\mathcal{D}\overline{q}\mathcal{D}q e^{\imath \int \mathrm{d}x^4     \mathcal{L}_q\left(\overline{q},q,\sigma,\phi\right)}
   \int \prod_a \mathcal{D}s_a \mathcal{D}p_a e^{\imath \int \mathrm{d}x^4 \mathcal{L}_r\left(\sigma,\phi,s_a,p_a\right)}
\end{align}
where $\mathcal{L}_q$ is now quadratic in the quark fields which enables its evaluation through a symmetry preserving heat kernel scheme (generalized to deal with non-degenerate current masses \cite{HKOsipovHiller}). The evaluation of $\mathcal{L}_r$ which is purely bosonic can be done integrating out the unphysical fields using the stationary phase approximation (SPA) which for the present analysis was done to leading order. Shifting the remaining mesonic fields, by demanding that their vacuum expectation is zero, quarks acquire a dynamical mass ($M_a=m_a+\Delta_a$) via the gap equations (here on the isotopic limit $m_u=m_d\neq m_s$) which must be solved together with the stationary phase condition
	\begin{align}
	\left\{
	\begin{array}[c]{rcc}
	h_u+\frac{N_c}{6 \pi^2}M_u\left(3 I_0 -  \left(M_u^2-M_s^2\right)I_1\right) = & 0	\nonumber\\
	h_s+\frac{N_c}{6 \pi^2}M_s\left(3 I_0 + 2\left(M_u^2-M_s^2\right)I_1\right) = & 0  \nonumber
	\end{array}
	\right.\nonumber\\
	\left\{
	\begin{array}[c]{rcc}
	 G h_u +\Delta_u +\frac{\kappa}{16}h_u h_s +\frac{g_1}{4}h_u\left(2 h_u^2+h_s^2\right)+\frac{g_2}{2}h_u^3= & 0	\nonumber\\
	 G h_s +\Delta_s +\frac{\kappa}{16}h_u^2 +\frac{g_1}{4}h_s\left(2 h_u^2+h_s^2\right)+\frac{g_2}{2}h_s^3= & 0  \nonumber
	\end{array}
	\right.
	\end{align}
	where $I_0$ and $I_1$ refer to quark loop integrals and the $h_a$ to the quark condensates.

\section{Results}

It has been shown that to insure global stability of the vacuum the following conditions must be fulfilled \cite{2}: $g_1>0$ , $g_1+3g_2>0$ and $G>\frac{1}{g_1}\left(\frac{\kappa}{16}\right)^2$. Thus the inclusion of the 't Hooft term without the 8-quark interaction leads to global instability (metastability in mean field approximation (MF))\cite{3,5} which can be seen in the analysis of the effective potential as a function of the quark condensate Fig. 1 (a, c in SPA, b in MF).

The model parameters (current quark masses, couplings and cutoff used in the regularization of the quark loop) are fixed using chosen scalar and pseudoscalar mesons masses and weak decay rates ($f_\pi$ and $f_K$) \cite{2}. 
Despite the dramatic effect on vaccum stability the meson spectra at $T=0$ remain basically unchanged when adding 8-quark interactions, due to the interplay of the parameters (mainly a decrease in $G$ when $g_1$ increases, accompanied by a decrease in the $\sigma$ meson mass). 

\begin{figure}[htp]
\begin{center}
\subfigure[4 q]{\includegraphics[width= 0.2 \linewidth]{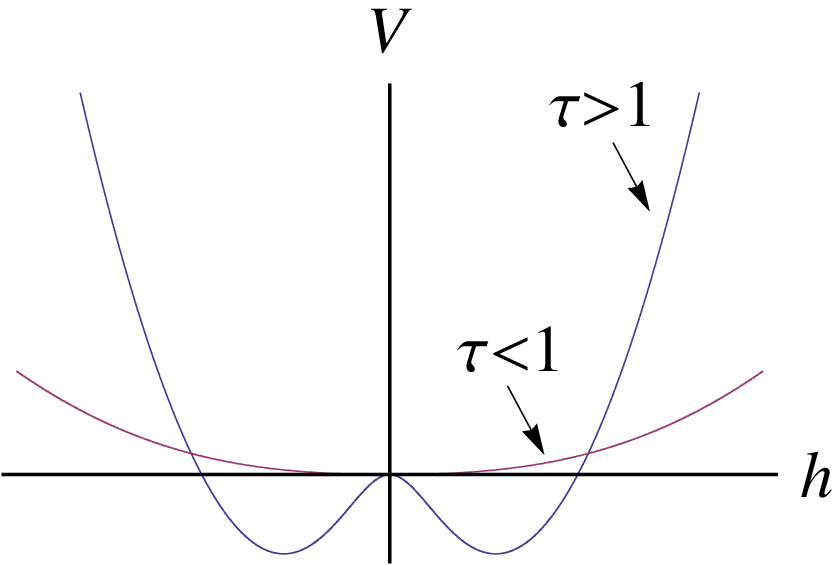}}
\subfigure[4 q + 6 q]{\includegraphics[width=0.2 \linewidth]{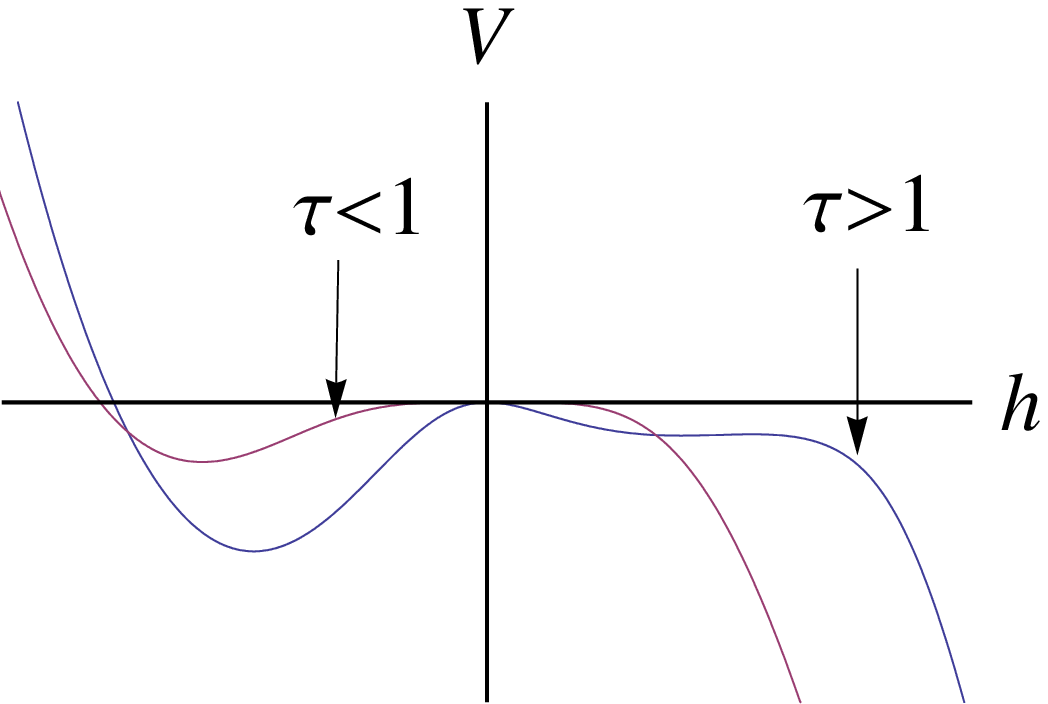}}
\subfigure[4 q + 6 q + 8 q]{\includegraphics[width=0.2 \linewidth]{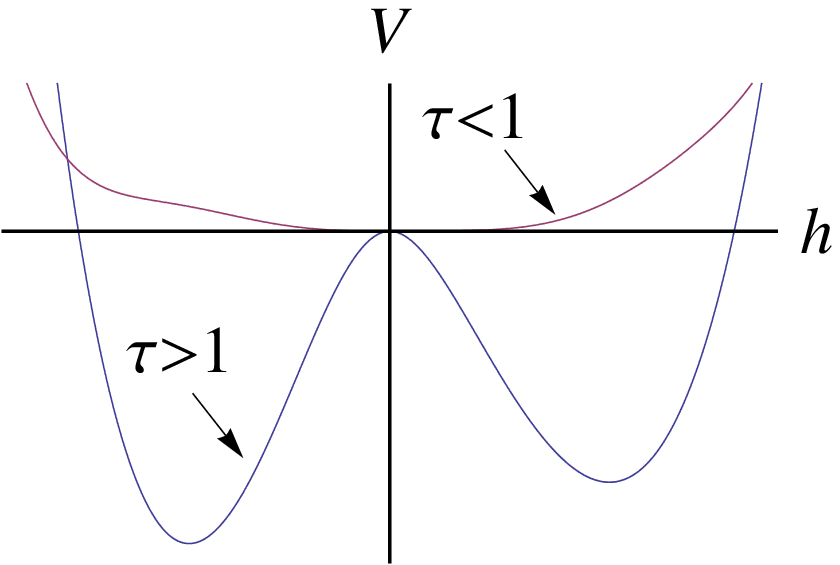}}
\caption{Shapes of the effective potential at $T=0$ for the cases with 4q, 4q + 6 q and 4q + 6q + 8q interactions as functions of the quark condensate in the $SU(3)$ chiral limit for different curvatures $\tau=Nc G \frac{\Lambda^2}{2\pi^2}$ at the origin.}
\end{center}
\label{grafs.eps}
\end{figure}

\begin{figure}[htp]
\begin{center}
\label{VeffectT.eps}
\subfigure[]{\includegraphics[width= 0.25 \linewidth]{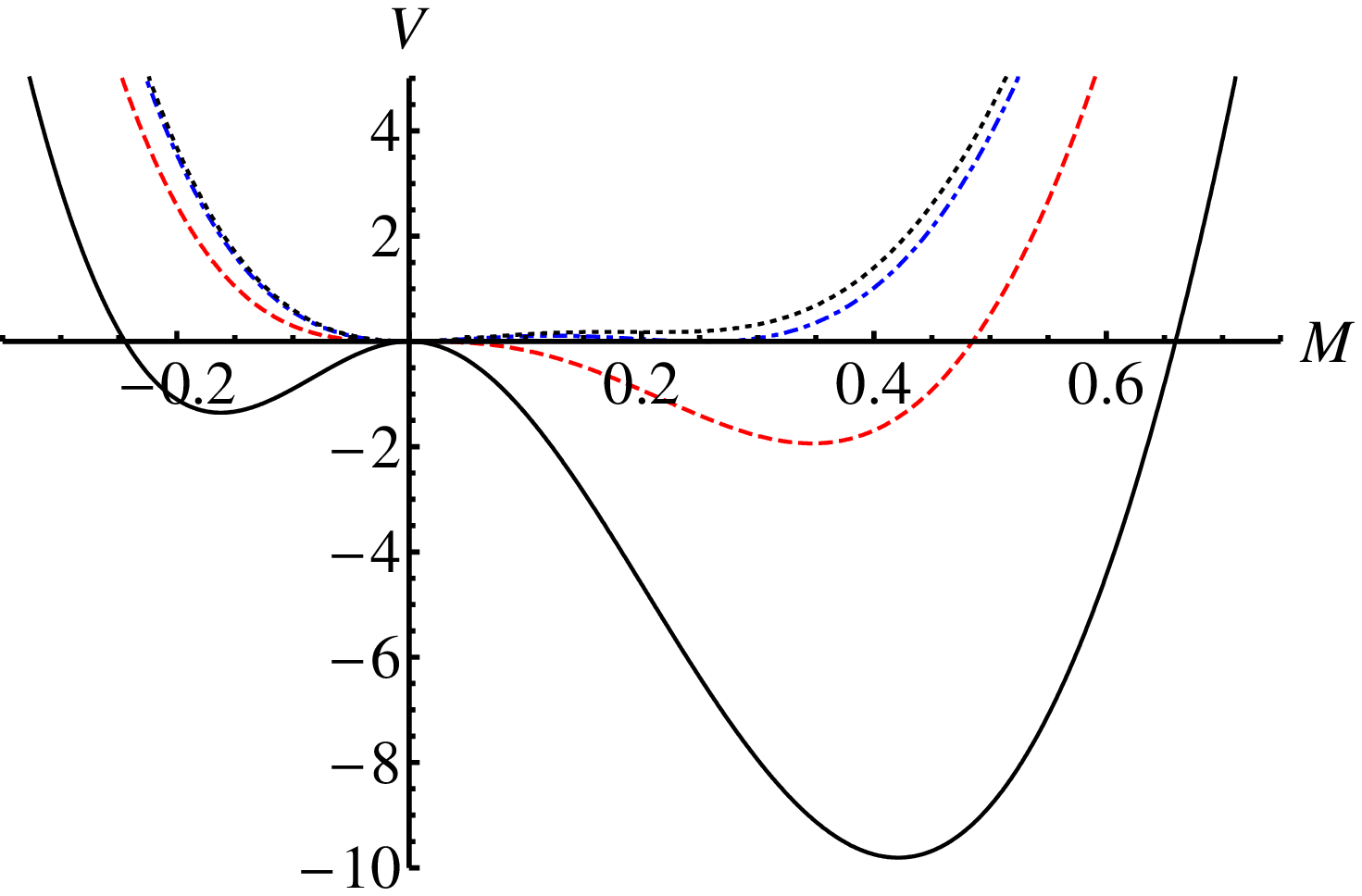}}
\subfigure[]{\includegraphics[width= 0.25 \linewidth]{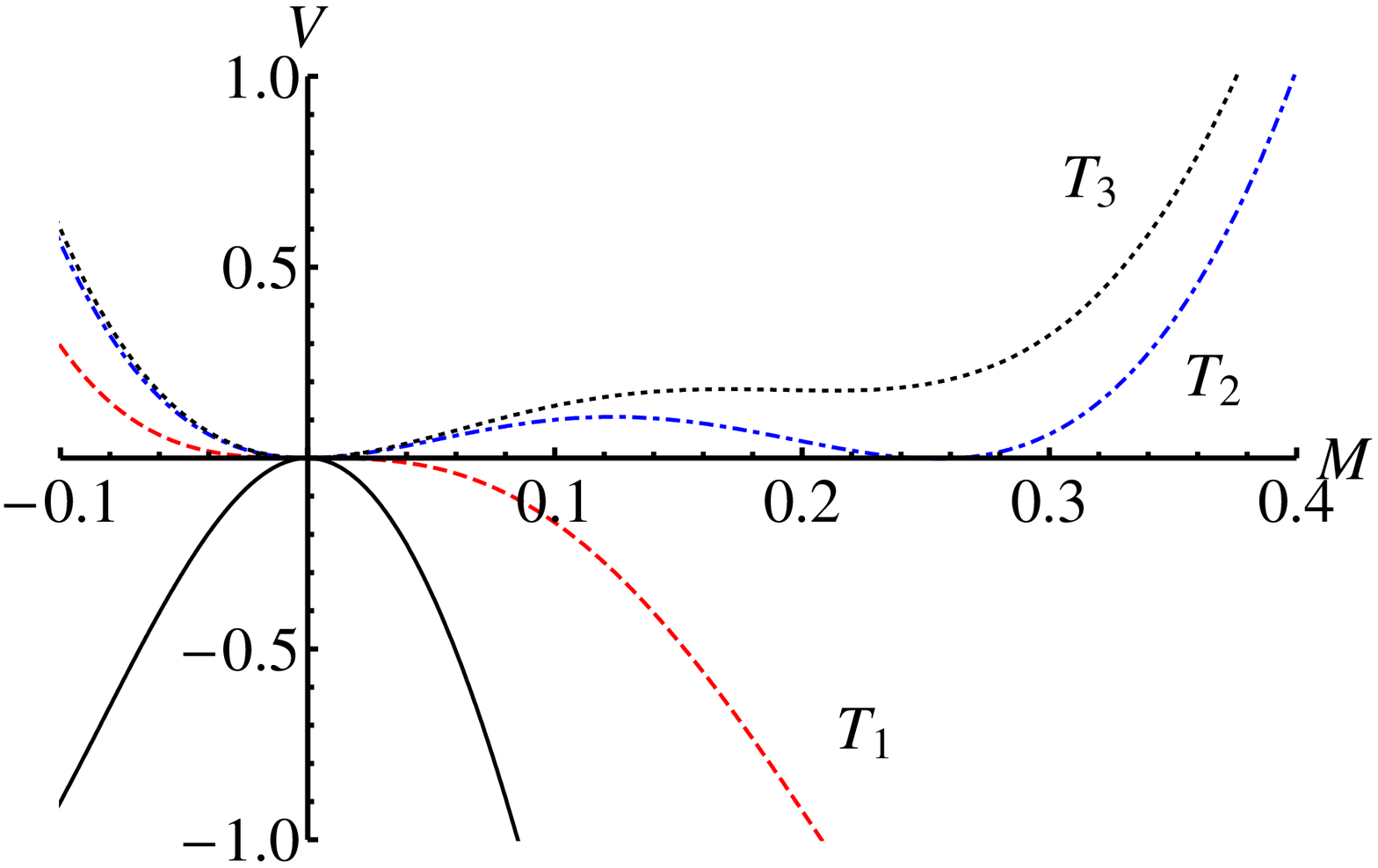}}
\caption{a) Effective potential (in $(10~\mathrm{GeV})^{-4}$) for several temperatures in the $SU(3)$ chiral limit shown as a function of the quark mass (in $\mathrm{GeV}$). The three temperatures $T_1$, $T_2$ and $T_3$ correspond to: appearance of a saddle point at the origin and after that a local mininum ($T_1$), degenerate two minima ($T_2$), and disappearance of the minimum corresponding to spontaneously broken symmetry ($T_3$). b) Close-up view of a).}
\end{center}
\end{figure}

At finite temperature the effect of 8 quark terms are however essential. The shape of the effective potential at finite $T$ runs through configurations which change the curvature at the origin (see Fig. 2). A significant decrease in the chiral transition temperature is achieved for an effective potential which at $T=0$ has curvature $\tau<1$ with coexistence of two vacua, a local mininum at the origin and a stable one induced by the 't Hooft six quark interaction strength.
In the case of realistic quark masses the minimum is shifted away from the origin.
Depending on the parameters we can get one or three sets of solutions $\{M^{(i)},M^{(i)}_s\}$ which correspond to the extrema of the thermodynamic potencial as can be seen in Fig. 3. 
If the OZI violating interaction strength, $g_1$, is greater than a critical value the three solutions are physical and when two of them meet there is a a first order transition  to the lower one; for lower values only one of the solution branches is physical and there is a rapid crossover when the other two cease to exist.
 
The temperature dependence of the meson mass spectra reflects the character of this transition as can be seen in Fig. 4 
 for the rapid crossover case \cite{4}.

\begin{figure}[htp]
\begin{center}
\label{massatemps.eps}
\subfigure[]{\includegraphics[width= 0.25 \linewidth]{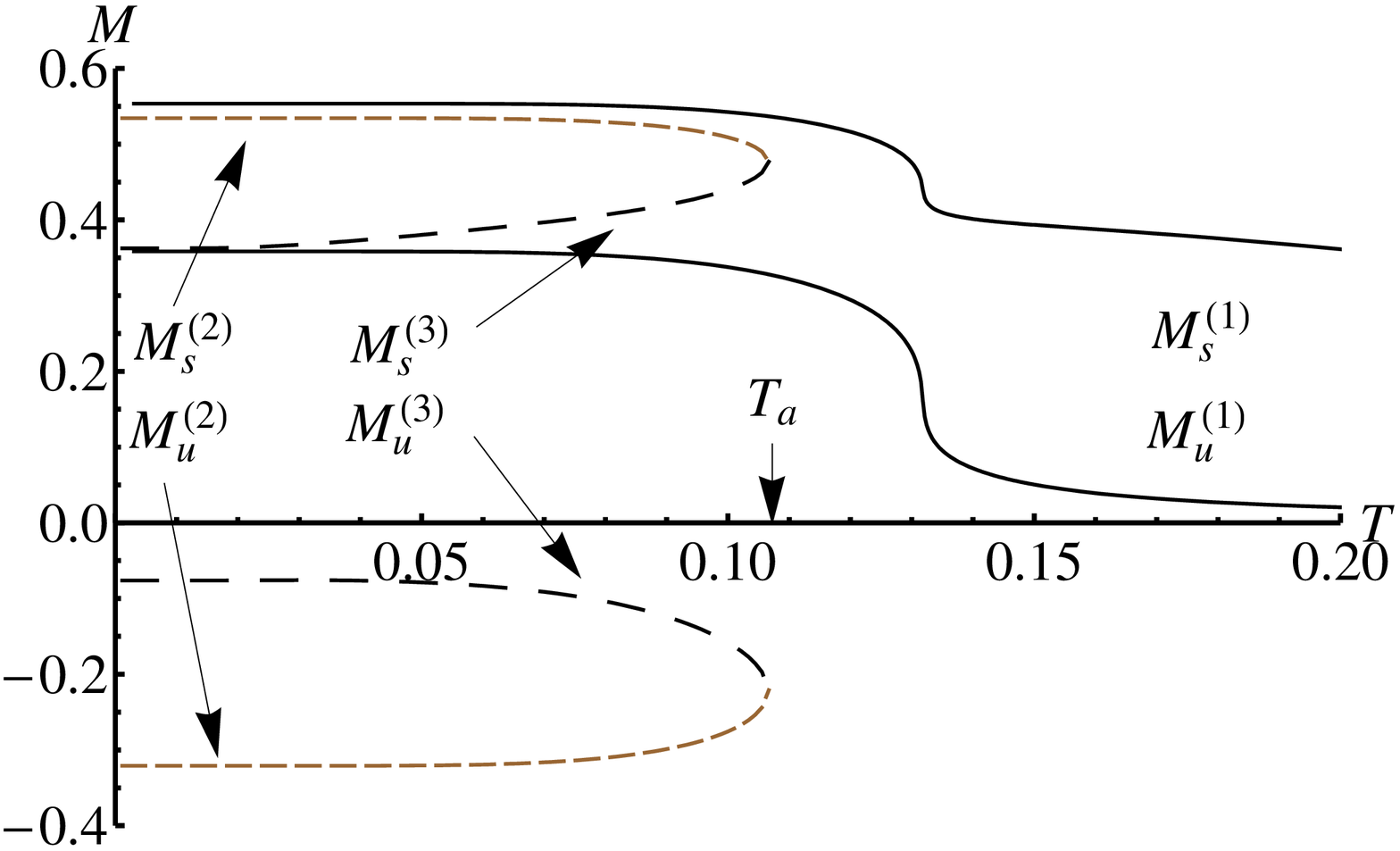}}
\subfigure[]{\includegraphics[width=0.25 \linewidth]{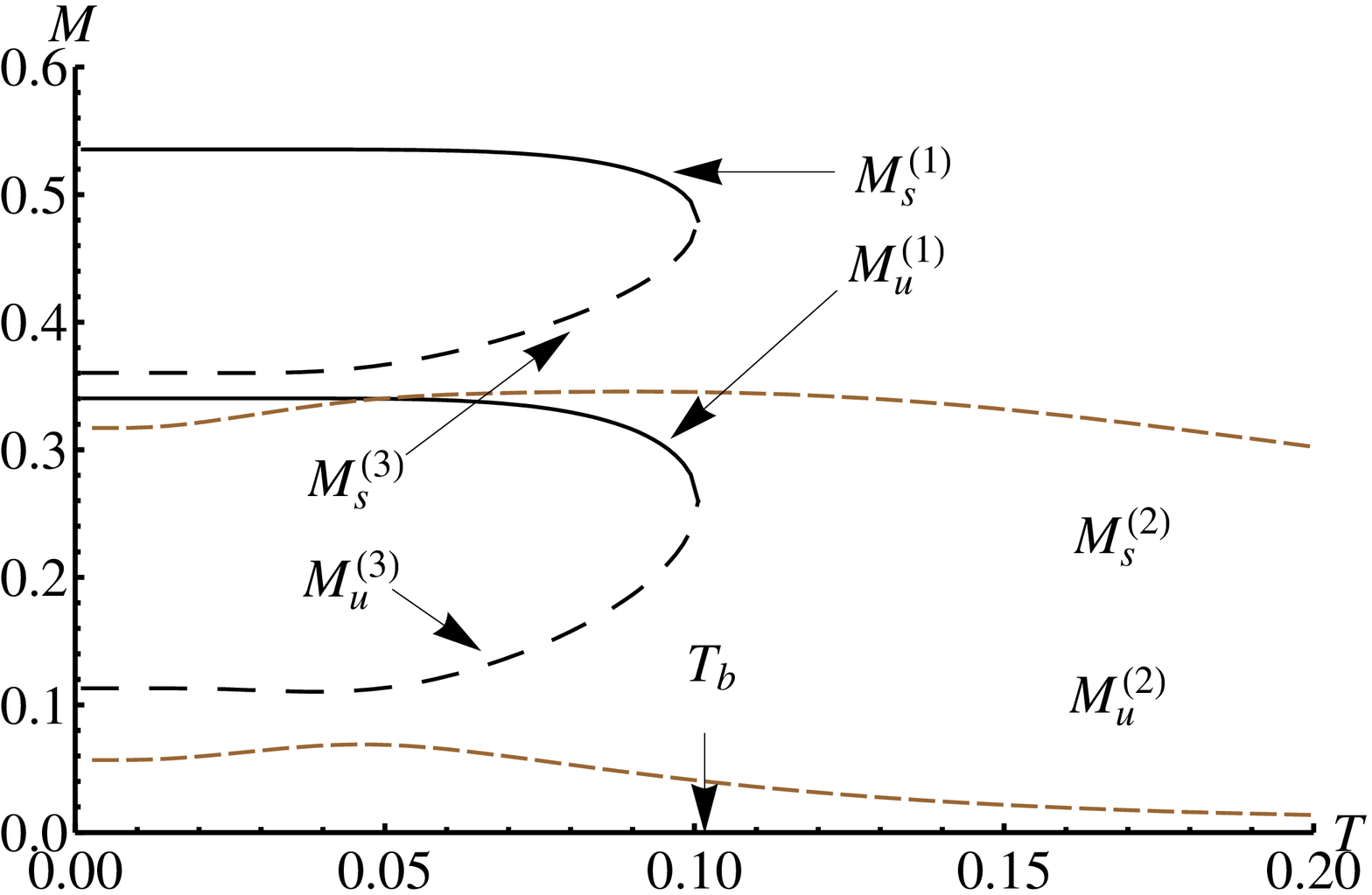}}
\caption{Quark mass solution pairs (in $\mathrm{GeV}$ ) (for details see \cite{4}): crossover case (a) and  the first order transition case (b).}
\end{center}
\end{figure}

\begin{figure}[htp]
\begin{center}
\label{massamesoes.eps}
\includegraphics[width= 0.5 \linewidth]{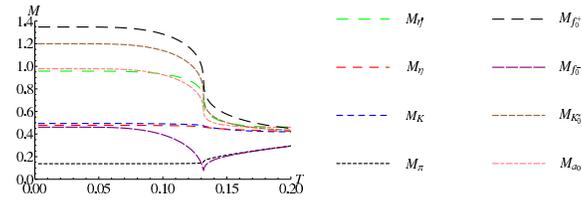}
\caption{Temperature dependence of the meson masses in the crossover case (all in units {$\mathrm{GeV}$}).}
\end{center}
\end{figure}

\section{Summary}
\noindent

The introduction of the eight quark interaction needed to ensure vacuum stability lowers the critical temperature for chiral transition while leaving the meson spectra at $T=0$ unaffected and introduces the possibility of first order or crossover transitions.They act as a chiral thermometer as it is their strength that determines the type of transition, its slope and the temperature at which it occurs.

This work has been partly supported by grants of Fundacao para a Ciencia e Tecnologia, POCI 2010 and FEDER, POCI/FP/63930/2005, POCI/FP/81926/2007 and SFRH/BD/13528/2003. \\

	

\end{document}